# Gravity and magnetic field dependent study of diffraction patterns in ferrofluids

M. Shalini (1), Chintamani Pai (2), S. Radha (2)

((1) UM-DAE Centre for Excellence in Basic Sciences, University of Mumbai, Mumbai-400098, (2) Department of Physics, University of Mumbai, Mumbai- 400098)

In this paper, we report the experimental observation of diffraction patterns in a ferrofluid comprising of $Fe_3O_4$ nanoparticles in hexane by a 10 mW He-Ne laser beam. An external dc magnetic field (0-2 kG) was applied perpendicular to the beam. The diffraction pattern showed variation at different depths of the sample in both zero and applied magnetic field. The patterns also exhibit a change in shape and size as the external field is varied. This effect is presumably arising due to thermally induced self-diffraction under the influence of gravity and external magnetic field.

## I. INTRODUCTION

Several studies report diffraction patterns from ferrofluids. In most of these cases, formation of these patterns are studied in the presence of an externally applied magnetic field. It is well known that magnetic field induces the formation of chain like structures along the field direction as observed in microscopic studies[1]. Many groups have observed diffraction patterns that corroborate chain formation[2-6]. On the other hand, some reports discuss the pattern obtained in a ferrofluid due to the agglomeration effect of the particles under a magnetic field. Bacri and Salin relate it to the effect arising due to the applied field forming an assembly of agglomerates having ellipsoidal shape[7]. The time-dependent evolution of such patterns probe the dynamics of structure formation in the presence of an external magnetic field[4,8]. Some researchers have studied the modulation of diffraction patterns as a function of varying magnetic field[9,10].

Apart from these, formation of such diffraction patterns have been reported in case of non-magnetic colloids and nematic liquid crystals as well in the absence of an external magnetic field[11-14]. In these cases, the patterns are attributed to the phenomena of thermal lensing coupled with self-diffraction. This happens when an intense Gaussian beam forms far-field patterns due to interaction with a thin self-defocusing material. This effect is known as spatial phase modulation due to intensity dependent complex refractive index in a non-linear material having contribution from thermal effects. Passing a Gaussian laser beam through a colloidal system creates local heating, due to which a refractive index gradient is established. Few groups have reported thermally induced self-diffraction patterns in magnetic fluids[15,16].

The motivation for our work follows the results reported by Shalini et al where transmission of light and its reflection from a commercially available ferrofluid has been studied experimentally[17]. In this paper, we report diffraction patterns from a ferrofluid consisting of 40 nm ferrite particles dispersed in hexane. We investigate evolution of thermally induced diffraction patterns under the influence of varying magnetic field as well as gravity.

## II. EXPERIMENTAL SETUP

The sample consisted of ferrite nanoparticles suspended in hexane as a dispersion medium. The nanoparticles were prepared by standard co-precipitation technique. The size of the particles as measured by dynamic light scattering (DLS) was 40 nm and the saturation magnetization measurements on a SQUID magnetometer (Quantum Design MPMS5) showed an $M_s$ of 40 emu/g and near zero coercivity at room temperature.

The experimental setup consists of an electromagnet capable of producing a variable magnetic field from 0 to 1.7 kG. The ferrofluid sample with a concentration of 35mg/ml was injected into a glass cuvette of path length 1mm and placed in between the pole pieces of the electromagnet. A 10 mW He-Ne laser of wavelength 633 nm was incident normally on the sample such that the beam falls at the centre of the cuvette. The schematic diagram of the experimental setup is shown in Fig.1. The direction of magnetic field is perpendicular to the direction of the incident light. The diffraction pattern produced was projected on a screen kept at a distance of 187 cm from the sample and having a grid of least count 1mm. These patterns were recorded by a charge-coupled device (CCD) camera.

Initially, depth profile of diffraction patterns is obtained by moving the cuvette vertically upward so that the laser beam is incident at different depths of the sample starting from the topmost layer to the bottom. In addition to this, diffraction patterns at given depth is recorded when magnetic field is varied between 0 to 1.7 kG. All readings were repeated in triplicate to ensure reproducibility.

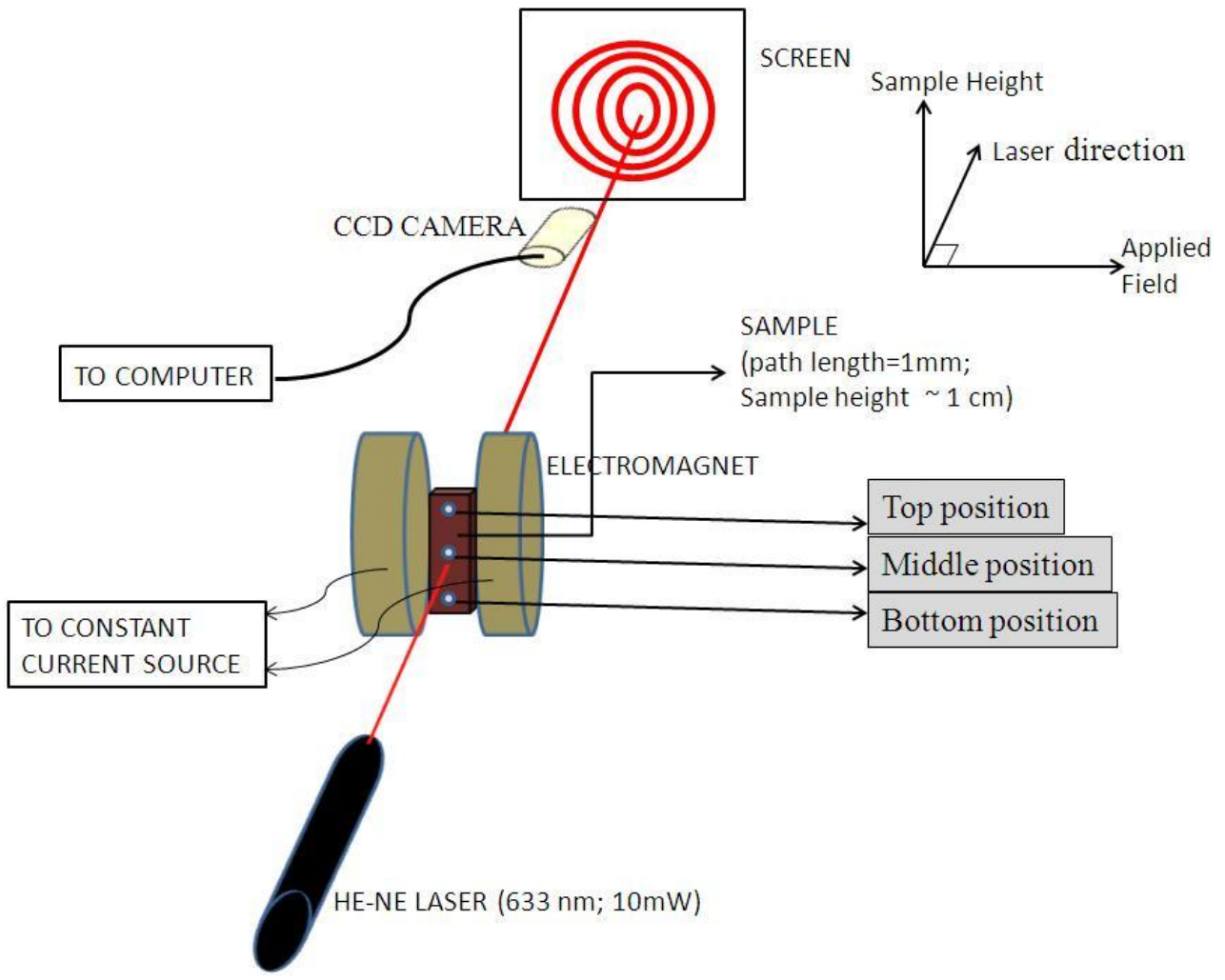

FIG. 1. The experimental set-up

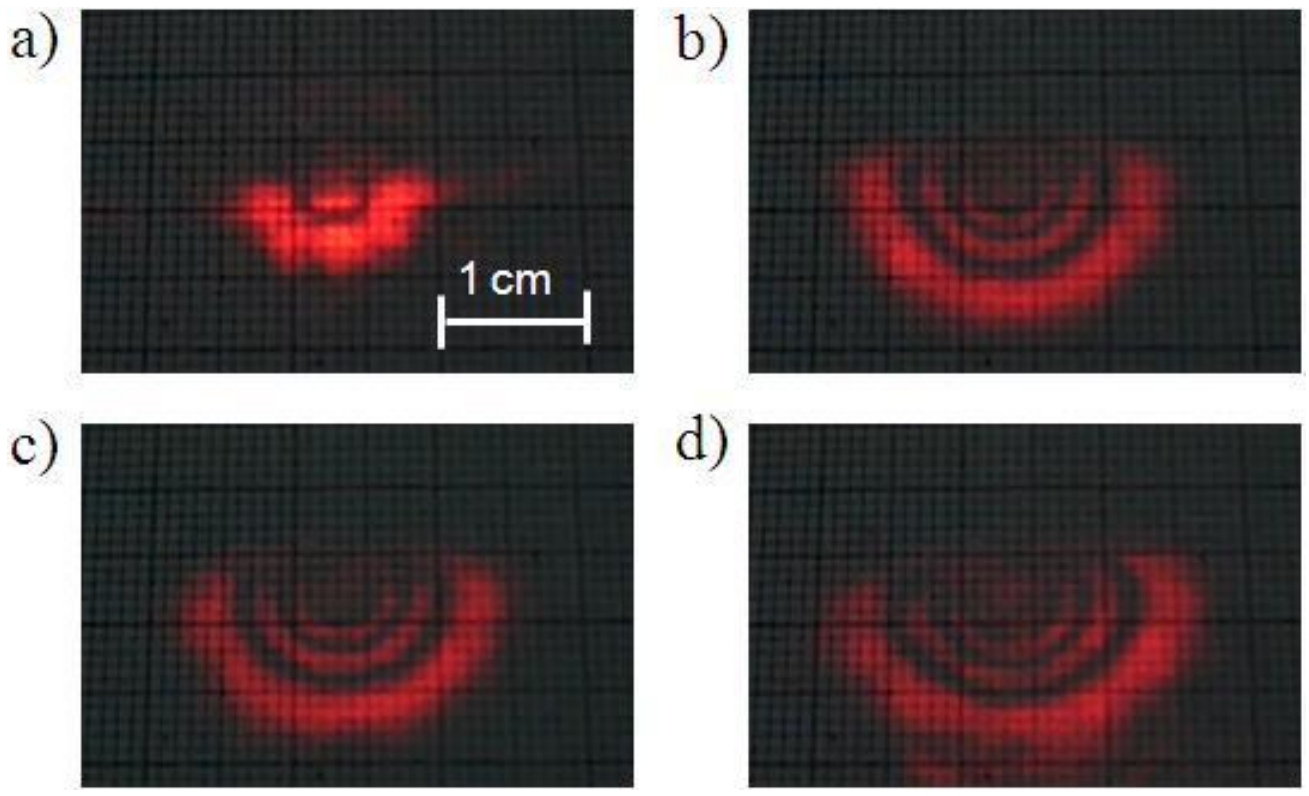

FIG.2. Diffractions patterns in the absence of magnetic field at a) topmost position b) 3mm below topmost position c) 6mm below topmost position d) bottom of the sample

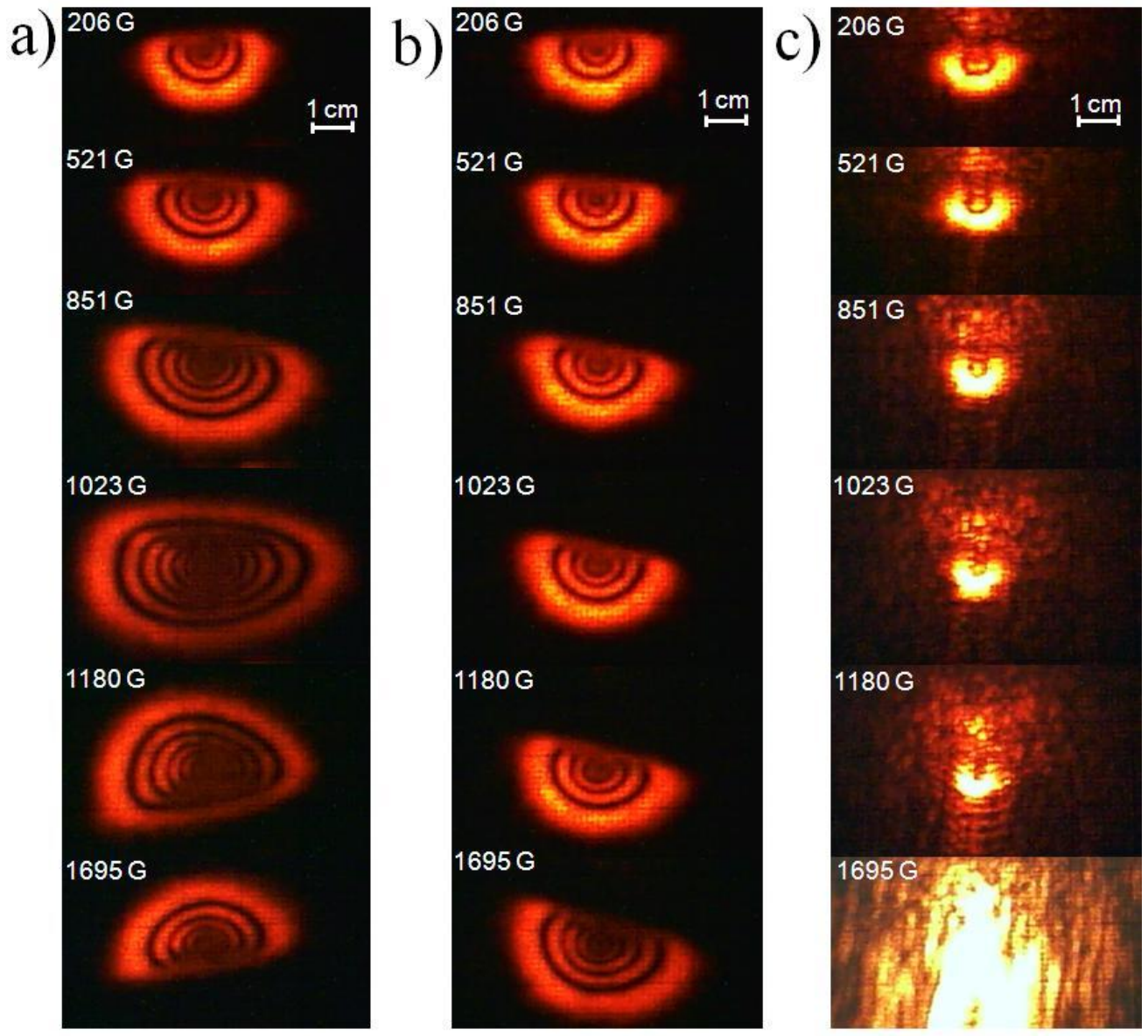

FIG.3. Variation in diffraction patterns in presence of magnetic field (206 G, 521 G, 851 G, 1023 G, 1180 G, 1695 G) at different positions in the sample along the vertical direction. a) Bottom b) Mid c) Top positions

## III. RESULTS AND DISCUSSION

Fig. 2 shows the diffraction patterns obtained at different positions along the height of the sample at zero field. As the laser beam is moved vertically down along the sample height, there is a gradual increase in the diameter of the diffraction rings. This seems to be due to the concentration gradient established within the sample. Due to gravity, particles settle down which results in increasing the concentration from top to bottom lay of the colloidal system. The diffraction patterns show a change with depth due to this concentration gradient. Researchers have reported gravitation dependent thermally induced self-diffraction patterns in carbon nanotube (CNT) solutions, where CNTs distribute themselves exponentially along the height[18]. As far as the shape of the rings is concerned, similar asymmetrical rings have been observed in CNT solutions with different laser power[11,18]. Our results clearly demonstrate the influence of gravity on the shape and size of the diffraction patterns.

Fig.3 shows the diffraction patterns from the bottom, middle and top of the sample as a function of applied magnetic field. It is clearly seen that the diffraction patterns evolve differently in these three regions with field.

At the bottom position, diffraction pattern grows in size and shows a gradual rotation with increase in the field. More number of rings emerge till the diffraction pattern becomes elliptical in shape at a particular value of the magnetic field (1023 G). After reaching an elliptical shape, with further increase in the field, it starts diminishing in size and the pattern shows an inversion about horizontal axis compared to a pattern at zero field. Except for the elliptical pattern at certain field, no complete rings are formed. For the patterns obtained at some point above the bottom position, a gradual rotation and change in size with increasing magnetic field is observed. However, there is no inversion or transformation into an elliptical pattern. At the top position, the pattern gets distorted when applied magnetic field is increased.

The observed complex evolution of diffraction patterns show the dynamics arising out of the interaction between gravity dependent concentration gradient, external magnetic field and thermally induced self diffraction. The chain formation in ferrofluids along the direction of the magnetic field itself can cause a change in the refractive index gradient established due to thermal lensing at given local concentration within the sample. Physical rotation and distortion in shape of agglomerates during the chain formation must be also taken into account. Observation of such physical rotation in thermally induced patterns is reported in case of nematic liquid crystals.[19] All these presumably lead to a complex dynamics involving rotation of asymmetrical patterns, as seen in our data. Further experiments on fluids with varying size and shape of the particles as well as the liquid medium and varying field geometry are required to understand their influence on the observed magneto-optic phenomena.

## IV. CONCLUSION

In this paper, we report an investigation of dynamic thermally induced diffraction patterns when coupled with the effect of gravity and external magnetic field, by transmitting a laser beam through a ferrofluid sample. Patterns show observable change due to change in position of the transmitted beam along the height of the sample and external magnetic field. The observation of diffraction patterns in zero field confirms the thermal lensing effect as evinced in case of non-magnetic colloids. The effect of gravity induced sedimentation and the field induced chain formation of the particles in the fluid contribute to the observed variation in the diffraction pattern. It is of further interest to study the dynamics of these patterns based on the direction of applied field, the nature of the particles as well as the carrier liquid. Such observations are expected to provide potential information for the design of magnetically tunable opto-fluidic sensors.

## ACKNOWLEDGEMENTS

The authors would like to thank UM-DAE CBS and Department of Physics, University of Mumbai, for the laboratory facilities provided. The authors express their sincere thanks to Prof. R.Nagarajan (UM-DAE CBS), Prof. Deepak Mathur (TIFR) and Dr. M.R. Press (UM) for valuable discussion and inputs. The authors also thank Prof. D. C. Kothari (Physics, UM), Dr. Jacinta D'Souza (UM-DAE CBS) and Dr.Varsha Kelkar (Biotech, UM) for providing facilities for sample preparation. We also extend our sincere thanks to Prof. A.K. Nigam (TIFR) for SQUID measurement**,** Prof. G. Krishnamurthy (TIFR) and Dr. S.B. Kulkarni (Institute of Science) for help in DLS.